\documentclass[10pt]{extarticle}

\usepackage[bordercolor=white,backgroundcolor=gray!30,linecolor=black,colorinlistoftodos, textsize=10pt]{todonotes}

\usepackage{amssymb}

\usepackage{textcomp}
\usepackage{amsmath}
\usepackage[textstyle,amssymb]{SIunits}
\usepackage{graphicx}
\usepackage{fixltx2e}  
\usepackage[normalem]{ulem}  

\usepackage[twocolumn,textwidth=183mm,columnsep=6mm,top=20mm,bottom=25mm]{geometry}
\usepackage{helvet}
\usepackage{titlesec}
\titleformat*{\section}{\bfseries\sffamily}
\titlespacing{\section}{0pt}{*4}{*0}
\titleformat{\subsection}[runin]{\normalfont\bfseries}{\thesubsection.}{3pt}{}
\usepackage{setspace}
\usepackage[breaklinks=true]{hyperref} 

\usepackage[font={footnotesize,sf},labelfont={sf,bf},labelsep=endash,justification=raggedright]{caption}

\usepackage{natbib}
\usepackage[labelfont=bf,justification=justified]{caption}

\begin{document}

\twocolumn[\begin{@twocolumnfalse}
	{\LARGE\sf \textbf{Synchronization of frequency combs by optical injection}}
	\vspace{0.5cm}
	
	{\sf\large \textbf {Johannes~Hillbrand$^{1,5}$, Mathieu~Bertrand$^{1}$, Valentin~Wittwer$^{2}$,  Nikola~Opa\v{c}ak$^{3}$, Filippos~Kapsalidis$^1$, Michele~Gianella$^4$, Lukas~Emmenegger$^4$, Benedikt Schwarz$^3$, Thomas~Südmeyer$^2$, Mattias~Beck$^1$, J\'er\^{o}me~Faist$^{1,6}$}}
		\vspace{0.5cm}
		
		{\sf \textbf{$^1$ Institute for Quantum Electronics, ETH Zurich, Auguste-Piccard-Hof 1, 8093 Zurich, Switzerland\\
				$^2$ Laboratoire Temps-Fr\'equence, Universit\'e de Neuch\^atel, 2000 Neuch\^atel, Switzerland\\
				$^3$ Institute of Solid State Electronics, TU Wien, Gusshausstrasse 25-25a, 1040 Vienna, Austria\\
				$^4$ Empa, Laboratory for Air Pollution / Environmental Technology, 8600 Dübendorf, Switzerland\\
				$^5$ e-mail: {jhillbra@phys.ethz.ch}\\
				$^6$ e-mail: {jfaist@ethz.ch}}}
		\vspace{0.5cm}
\end{@twocolumnfalse}]
\vspace{0.5cm}

{\sf \small \textbf{\boldmath
		\noindent Optical frequency combs based on semiconductor lasers are a promising technology for monolithic integration of dual-comb spectrometers. However, the stabilization of offset frequency f$_{ceo}$ remains a challenging feat due the lack of octave-spanning spectra. In a dual-comb configuration, the uncorrelated jitter of the offset frequencies leads to a non-periodic signal resulting in broadened beatnotes with a limited signal-to-noise ratio (SNR). Hence, expensive data acquisition schemes and complex signal processing are currently required. Here, we show that the offset frequencies of two frequency combs can be synchronized by optical injection locking, which allows full phase-stabilization when combined with electrical injection. A single comb line isolated via an optical Vernier filter serves as Master oscillator for injection locking. The resulting dual-comb signal is periodic and stable over thousands of periods. This enables coherent averaging using analog electronics, which increases the SNR and reduces the data size by one and three orders of magnitude, respectively. The presented method will enable fully phase-stabilized dual-comb spectrometers by leveraging on integrated optical filters and provides access for measuring and stabilizing $f_{ceo}$.
	}
}

\section{Introduction}

The spectrum of an optical frequency comb is given by a series of evenly spaced lines\cite{udem2002optical}. The frequencies $f_n$ of these modes are determined by merely two parameters: the repetition frequency $f_{rep}$, which equals the cavity roundtrip frequency and sets the line spacing, and the carrier envelope offset frequency $f_{ceo}$.
\begin{equation}
    f_n=n\cdot f_{rep}+f_{ceo} \quad \mathrm{with} \quad n \in \mathbb{N}
\end{equation}
Since both $f_{rep}$ and $f_{ceo}$ are usually located in the radio-frequency (RF) domain, frequency combs act as coherent gears linking optical frequencies to state-of-the-art RF electronics. The first frequency combs leveraged on this property to push the limits of precision metrology\cite{udem1999absolute}. More recently, their potential for optical sensing applications is gaining significant traction. In particular, dual-comb spectroscopy\cite{keilmann2004time,coddington2016dual} holds great promises for multi-species molecular sensing. In this technique, two frequency combs with a slightly different mode spacing are combined on a high-speed photodetector. The latter detects the multiheterodyne beating $S(t)$ between adjacent comb lines, which is called dual-comb interferogram. This signal is given by
\begin{equation}
    S(t)=\sum_n A_n B_n^* e^{i(n\Delta f_{rep}+\Delta f_{ceo})t}+c.c. \label{eq:dualcomb}
\end{equation}
where $A_n$ and $B_n$ are the modal electric field amplitudes of the two combs. Hence, the two optical spectra are coherently mapped to an electronic signal. Provided that $\Delta f_{rep}$ and $\Delta f_{ceo}$ are both constant in time, the dual-comb amplitudes $|A_nB_n^*|$ are simply retrieved by applying a Fourier transform. This allows to extract the absorption spectrum of a sample in the beam path. The temporal resolution of the dual-comb measurement is set by $\Delta f_{rep}^{-1}$, ranging from tens of nanoseconds to microseconds.

\begin{figure*}[h!]
\centering\includegraphics[width=1\textwidth]{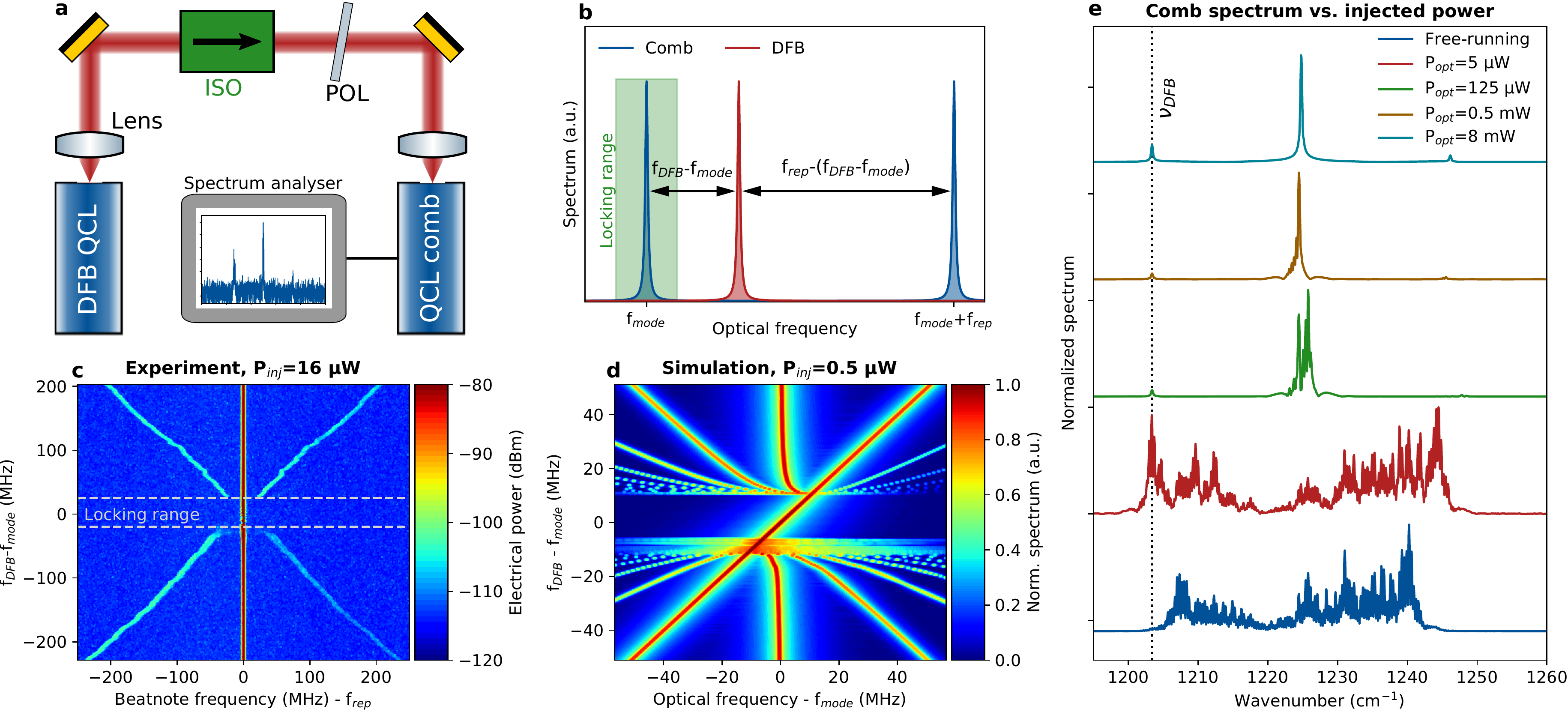}
\caption{\textbf{a}: optical injection locking of a QCL frequency comb to a single-mode DFB QCL. ISO: optical isolator. POL: polarizer for optical attenuation. \textbf{b}: the DFB QCL beats with its neighboring comb lines at $f_{DFB}-f_{mode}$ and $f_{rep}-(f_{DFB}-f_{mode})$. When the frequency of the injected radiation is within the locking range, the frequency comb is locked and only one beating at $f_{rep}$ exists. \textbf{c}: RF beatnote map of the QCL frequency comb, while the frequency of the DFB QCL (corresponding to the weak diagonal lines) is swept across a comb mode. Only the comb beatnote at $f_{rep} \approx $9.6~GHz is visible within a locking range of roughly 45 MHz. \textbf{d}: locking of a comb mode simulated in the optical domain using the Maxwell-Bloch equations. Again, the diagonal line represents the frequency of the injected signal. Merely 500 nW of injected power lead to a 20 MHz wide locking range. \textbf{e}: spectra of the QCL frequency comb depending on the injected power. The grey dashed line denotes the frequency of the DFB QCL.}\label{fig:fig1}
\end{figure*}

Naturally, the mid-infrared (mid-IR) region is of great importance, because it contains the fundamental roto-\\vibrational absorption lines of most molecules (molecular fingerprint). The development of mid-IR frequency combs based on semiconductor lasers\cite{hugi2012mid,schwarz2019monolithic,bagheri2018passively} is particularly appealing, because they are microchip-sized and electrically pumped. Dual-comb spectroscopy based on quantum cascade laser (QCL) frequency combs\cite{villares2014dual,wang2014high} was successfully demonstrated and used for monitoring protein reactions\cite{klocke2018single}, combustion analysis\cite{pinkowski2020dual} and hyperspectral imaging\cite{sterczewski2019terahertz}. Additionally to the sub-\textmu s temporal resolution, spectral interleaving enables a spectral resolution of at least 0.001 cm$^{-1}\,$\cite{gianella2020high}. Furthermore, the high optical output power of QCLs up to the Watt-level combined with a large mode spacing on the order of 10 GHz results in an unrivalled brightness with mW-level power per mode\cite{jouy2017dual}. However, the aforementioned advantages come at a price. Both $f_{rep}$ and $f_{ceo}$ of QCL frequency combs depend sensitively on environmental conditions such as temperature and driving current\cite{hillbrand2018tunable}. As a consequence, the signal in Eq. \ref{eq:dualcomb} is non-periodic and varies over time, resulting in a broadening of the dual-comb lines. This fact increases the complexity of data acquisition and signal processing considerably. Hence, stabilization techniques for both $\Delta f_{rep}$ and $\Delta f_{ceo}$ is expected to significantly reduce the complexity and thus cost of QCL dual-comb spectroscopy, while increasing their sensitivity.

In this work, we show that $f_{rep}$ and $f_{ceo}$ of two QCL frequency combs can be fully phase-locked to each other by combining electrical and optical injection locking. Both repetition frequencies are readily stabilized by electrical injection locking\cite{hillbrand2019coherent}. The offset frequencies are synchronized to each other via optical injection locking enabled by a passive optical filter based on the Vernier effect. The resulting dual-comb signal is not only stable over thousands of periods, but also harmonic (i.e., $\Delta f_{ceo}=0$), which leads to a periodic waveform. This allows coherent averaging of the interferogram using analog electronics and reduces the amount of data by more than three orders of magnitude, while retaining all spectral information.

\section{Optical injection locking}

In order to stabilize the dual-comb interferogram in Eq. \ref{eq:dualcomb}, all degrees of freedom of both combs have to be locked to each other. Both repetition frequencies can be controlled by injecting the signal of an external RF oscillator, whose frequency is close to the mode spacing, into the laser cavity\cite{hillbrand2019coherent,stjean2014injection,forrer2019photon}. However, the control of $f_{ceo}$ is more challenging. Since QCL frequency combs with octave-spanning spectra still remain an elusive goal, absolute frequency stabilization via f-2f interferometry\cite{telle1999carrier,jones2000carrier} is not yet possible. However, according to Eq. \ref{eq:dualcomb} it suffices to lock both offset frequencies relative to each other to achieve a stable interferogram. Multiple strategies based on digital post-processing\cite{burghoff2016computational,sterczewski2019computational} and feedback loops\cite{consolino2019fully,komagata2021coherently} have been proposed to reach this goal. Here, we choose a different strategy and stabilize $\Delta f_{ceo}$ via optical injection locking. The frequency and phase of a slave laser can be locked to a Master laser by injecting its light into the slave cavity\cite{pantell1965laser,tang1967phase}. Hence, locking a single line of both combs to the same Master laser allows to synchronize the offset frequencies of both combs\cite{vangasse2020chip}.

In order to investigate the impact of optical injection, the single-mode radiation emitted by a distributed-feedback (DFB) QCL was injected into a QCL frequency comb (Fig. \ref{fig:fig1}a). Several beatnotes can be extracted directly from the driving current of the frequency comb (Fig. \ref{fig:fig1}b) thanks to the fast carrier dynamics of QCLs. Firstly, the equidistant modes of the comb beat with each other resulting in the comb beatnote at $f_{rep}$. Due to the high coherence between the comb lines, this beatnote is extremely narrow with a linewidth below 1 kHz. Secondly, the injected signal of the DFB QCL produces two beatings with its neighboring comb modes, one being at $f_{DFB}-f_{mode}$ and the other at $f_{rep}-(f_{DFB}-f_{mode})$. The linewidth of these beatings is on the MHz level and thus considerably larger than the comb beatnote, because the phases of the DFB QCL and the frequency comb are uncorrelated while unlocked. However, if the optical frequency of the DFB QCL is within the locking range of a comb mode, the latter is phase-locked to the DFB QCL. Under these conditions, the linewidth of the DFB beatnote collapses to the narrow width of the comb beatnote. Fig. \ref{fig:fig1}c shows the RF spectrum extracted from the QCL frequency comb around $f_{rep}$ while the DFB frequency is swept across a comb mode. The strong and narrow line in the center is the intrinsic intermode beatnote, whereas the diagonal lines represent the swept frequency of the DFB QCL.  Within a locking range of roughly 45 MHz, the beating between the DFB and the comb mode collapses and only the narrow intermode beatnote remains, which indicates successful phase-locking. The magnitude of the locking range is remarkable, considering the small injected power of merely 16 \textmu W (excluding coupling losses). This process can be modelled directly in the optical domain using the Maxwell-Bloch equations\cite{opacak2019theory}. The simulation results are shown in Fig. \ref{fig:fig1}d and reveal a locking range of approximately 20 MHz for 500 nW of injected optical power. The diagonal line represents the injected signal, whose frequency is swept, and the vertical line the comb mode. It should be noted that the locking behaviour depends strongly on the injected power and the linewidth enhancement factor of the comb line\cite{opacak2021spectrally,opacak2021frequency}. Depending on these parameters, instabilities are observed, which depend critically on the frequency tuning direction and can either enhance or decrease the locking range considerably (see supplementary section 1).

Optical injection locking was previously investigated for enforcing single-mode operation of high-power, but multi-mode lasers\cite{winkelmann2011injection}. Therefore, it is expected that the injected radiation has a strong impact on the comb spectrum. The latter is shown in Fig. \ref{fig:fig1}e as function of the injected power. The spectrum broadens symmetrically at its red and blue edge upon injecting only 5 \textmu W of optical power. The new edge of the spectrum is determined by the frequency of the DFB QCL. As the injected power is further increased, the spectral width decreases considerably and finally consists of only one dominant mode. Therefore, around 10 \textmu W are sufficient to lock the comb, while higher power results in the collapse of the comb spectrum.

\begin{figure}[h!]
\centering\includegraphics[width=0.5\textwidth]{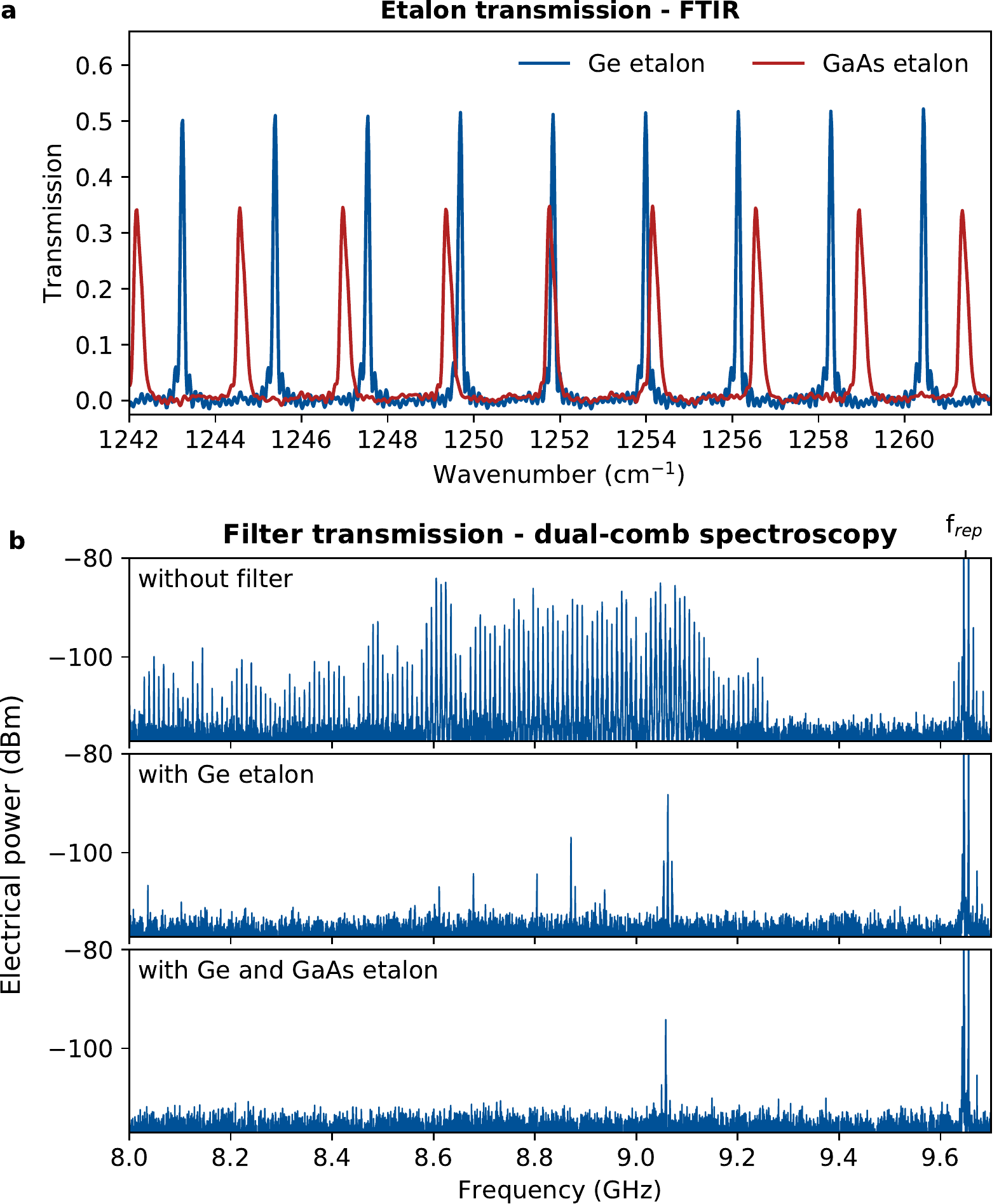}
\caption{\textbf{a}: FTIR transmission spectrum of two etalon filters based on Germanium and GaAs. The slightly different spacing of the resonances allows to isolate single comb lines via the Vernier effect. A more accurate measurement of the resonance width using a DFB QCL can be found in supplementary section 2. The fabrication details are in the Methods section. \textbf{b}: beating extracted from a QCL frequency comb, when two combs are shined into each other with no filter in the beam path (top), one etalon (middle) and both etalons acting as Vernier pair (bottom).}\label{fig:fig2}
\end{figure}

The individual modes of high-performance QCL frequency combs have a power on the mW level\cite{jouy2017dual}. Hence, a single line of one comb has sufficient power to act as Master oscillator for locking the second comb (setup schematic in supplementary section 3). This method eliminates the need for an additional DFB QCL. In order to isolate a single comb line, we employ a passive optical filter based on two Fabry-P\'erot interferometers (etalons). The transmission spectrum of both etalons was measured using a Fourier transform infrared (FTIR) spectrometer (Fig. \ref{fig:fig2}a) and consists of regularly spaced resonances with a peak transmission between 35-50~\%. Furthermore, the free spectral ranges of both etalons are slightly different. Hence, any desired mode of the comb spectrum can be isolated via the Vernier effect by tuning both etalons on resonance (e.g. by rotation). This tunability of the optical Vernier filter is particularly important. Since the two transmitted modes are locked to each other and thus have the same frequency, their beat frequency is zero. The neighboring lines, on either side of the locked mode, produce a beatnote at the same frequency equal to $\Delta f_{rep}$, $2\cdot\Delta f_{rep}$ and so on. Hence, the dual-comb spectrum will be folded around the locked comb line. Locking a mode at the edge of the spectrum therefore maximizes the optical bandwidth usable for spectroscopy. 


\begin{figure*}[h!]
\centering\includegraphics[width=1\textwidth]{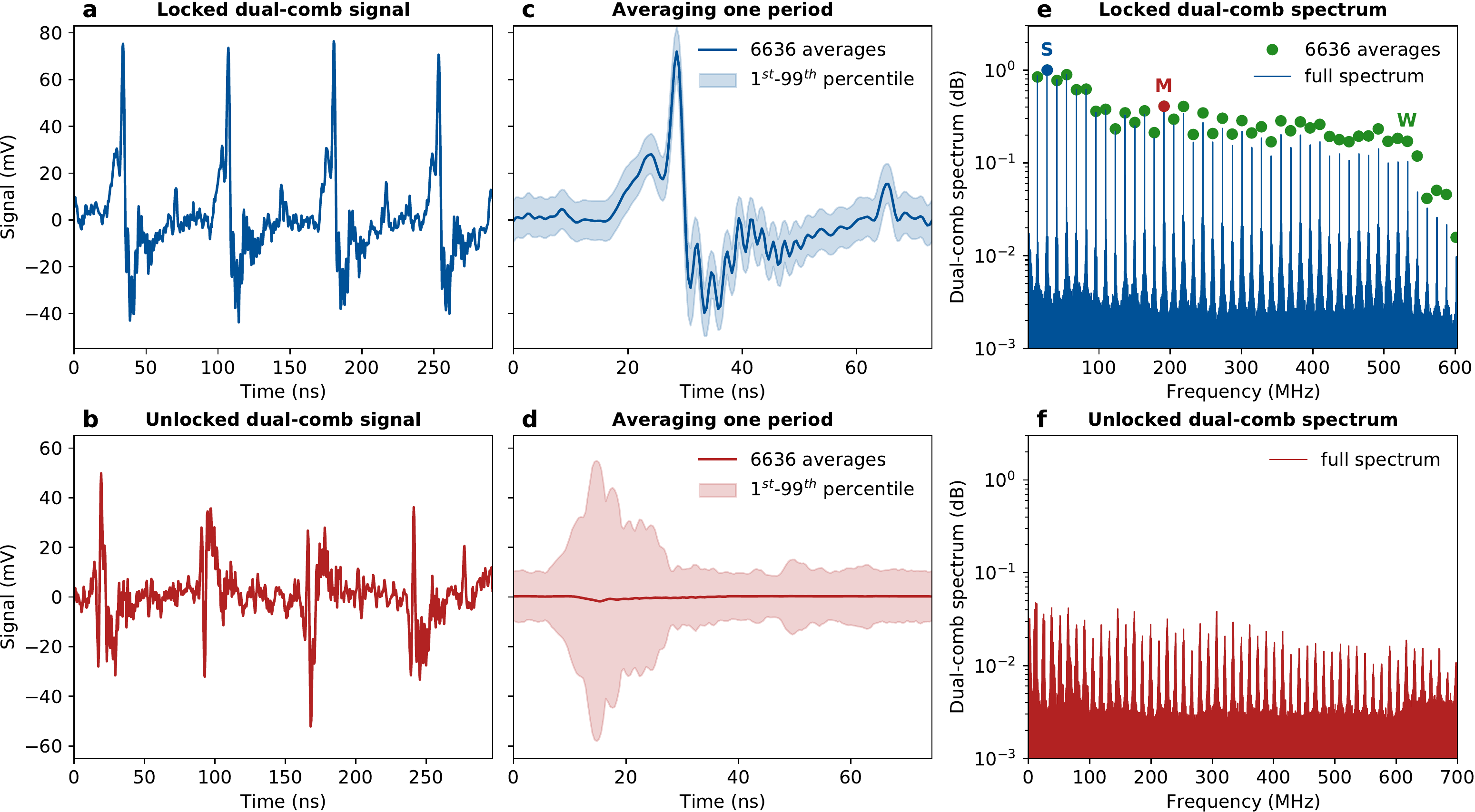}
\caption{\textbf{a}: dual-comb signal of two injection locked QCL frequency combs, leading to a periodic series of bursts. \textbf{b}: single period of the signal in \textbf{a} averaged over 6636 times. The shaded area represents the 1$^{\mathrm{st}}$ to 99$^{\mathrm{th}}$ percentile of all data points. The waveform remains almost identical to the original signal. The coherence of the averaging process is shown digitally here. Supplementary section 4 shows that this process is also possible using the averaging function of an oscilloscope.  \textbf{c}: dual-comb spectrum of the full interferogram (blue line) and one averaged period (green dots). \textbf{d}: dual-comb signal of two unlocked QCL frequency combs. A phase-slip is clearly visible between consecutive periods. \textbf{e}: averaged dual-comb interferogram without locking. The amplitude decreases drastically due to the non-periodic signal. \textbf{f}: unlocked dual-comb spectrum with a decreased SNR and broadened lines due to the f$_{ceo}$ jitter.}\label{fig:fig3}
\end{figure*}

The working principle of the optical Vernier filter is demonstrated in Fig. \ref{fig:fig2}b. The two QCL frequency combs are shined directly into each other. Without requiring an external photodetector, the dual-comb beating is extracted from the laser current via a bias-T. When no etalon is in the beam path, the full dual-comb spectrum is visible. Upon inserting the Ge etalon, one line is predominantly transmitted. However, several other modes are also partially transmitted due to the relatively broad resonances of a single etalon. As soon as both etalons are inserted into the beam path at the correct angle to transmit the desired mode, only a single line is isolated. Under these conditions, both combs can be coherently locked to each other by adjusting the driving current of one QCL, such that the comb line of that laser is within the locking range around the transmitted comb line of the other laser.

\section{Coherent dual-comb spectroscopy}

The injection locking method can be integrated into any dual-comb spectrometer using simply two beamsplitters (see supplementary section 3 for detailed setup). As a consequence of the mutual phase-lock between the combs, the dual-comb spectrum is expected to consist of extremely sharp lines separated by $\Delta f_{rep}$. Furthermore, since one line of both combs is locked to the same frequency, $\Delta f_{ceo}$ in Eq. \ref{eq:dualcomb} vanishes. Thus, the dual-comb interferogram is periodic, which is shown in Fig. \ref{fig:fig3}a. The signal consists of a series of consecutive bursts without any phase-slip occurring between them. 

This periodicity enables direct coherent averaging without requiring post-processing. To showcase this capability, the full time trace containing more than 6600 periods was sliced periodically at multiples of ${\Delta f_{rep}}^{-1}$. These slices were then aligned using the peak of the bursts and averaged, which simulates the triggered acquisition of an oscilloscope. Alternatively, $\Delta f_{rep}$ can be obtained by mixing the two RF signals used for locking both repetition frequencies and subsequently used as trigger source. The averaged signal shows the same waveform as the original interferogram without any loss of amplitude, which confirms coherent averaging.

Despite a reduction of the data size by over three orders of magnitude, the averaged signal contains the same spectral information as the full interferogram (Fig. \ref{fig:fig3}c). The full dual-comb spectrum shows numerous narrow lines at harmonics of $\Delta f_{rep}$, whose width is limited by the acquisition time over the full bandwidth (2 kHz). Since the dual-comb signal is harmonic, the Fourier transform of the averaged signal aligns exactly with the peak amplitudes of the full spectrum. Without optical injection, the dual-comb signal still shows a series of bursts (Fig. \ref{fig:fig3}d). However, a phase-slip is clearly visible between consecutive periods, resulting in a non-periodic waveform. Under these conditions, coherent averaging is not possible, as shown in Fig. \ref{fig:fig3}e. As a result, the lines of the dual-comb spectrum are broadened to roughly 1 MHz, while their signal-to-noise ratio is decreased by more than an order of magnitude.

\begin{figure}[h!]
\centering\includegraphics[width=0.5\textwidth]{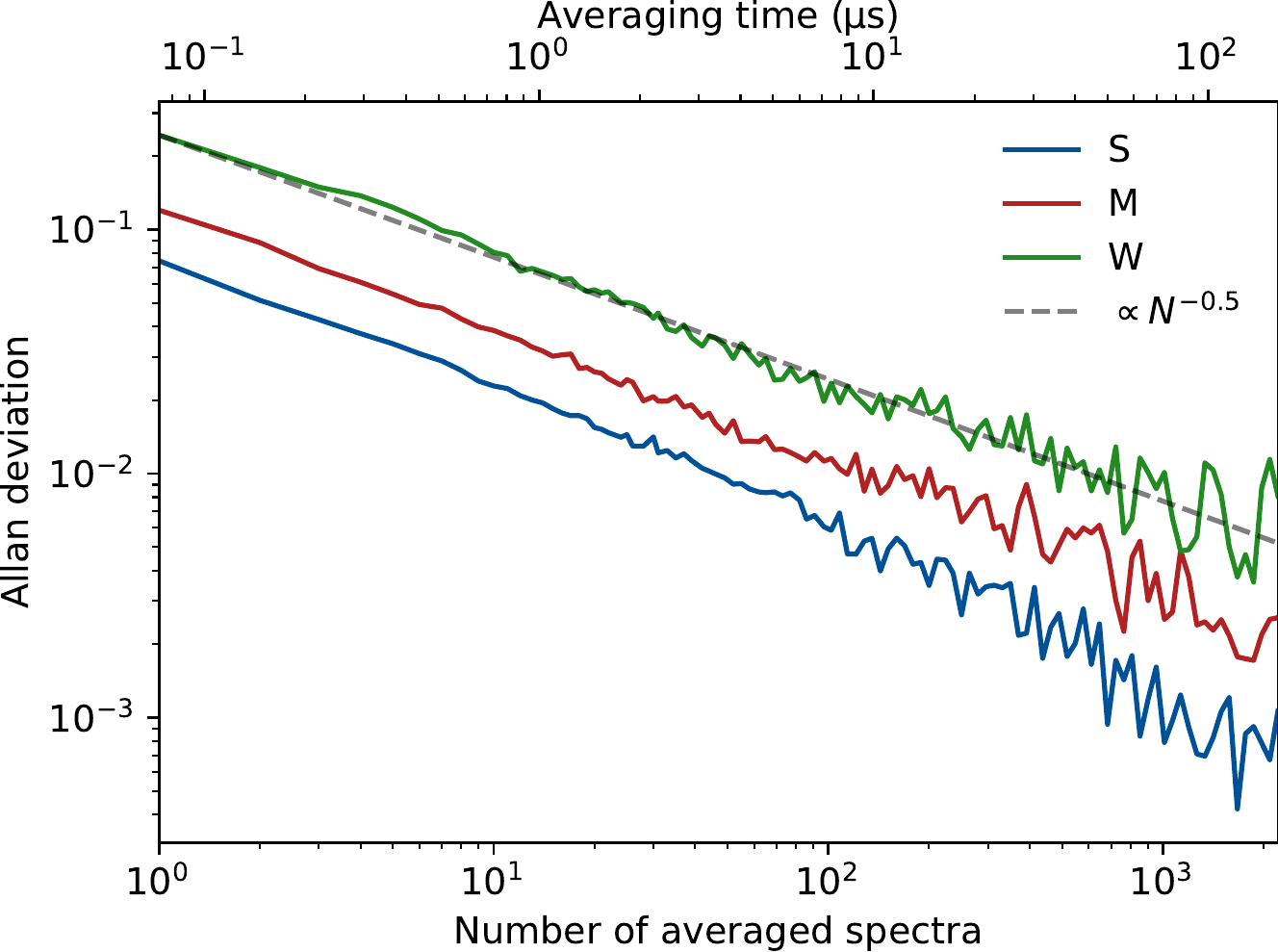}
\caption{Amplitude Allan deviation of three representative lines of the locked dual-comb spectrum (marked with S, M and W in Fig. \ref{fig:fig3}c). It should be noted that the Allan deviation reaches a local minimum for longer acquisition times (supplementary section 5). We attribute this to vibrations in the optical path, which also lead to phase variations of the dual-comb beatnotes.}\label{fig:fig4}
\end{figure}

The amplitude Allan deviation of three representative modes (Fig. \ref{fig:fig4}) further confirms the coherence of the locked dual-comb spectrum. As expected for a coherently averaged spectrum, all three lines decrease with the square root of the number of averaged periods. It should be noted that the Allan deviation reaches a minimum at an averaging time around 1~ms and decreases again above 10 ms (see supplementary section 5). However, the phase-lock between the two frequency combs is stable for at least several minutes. Likewise, the dual-comb signal can be coherently averaged for more than 1 million periods. For this reason, we attribute the local minimum of the Allan deviation to other drifts in our setup, such as mechanical vibrations and thermal expansion.

\section{Conclusion}

Our results demonstrate that QCL frequency combs can be coherently locked to an external Master oscillator via optical injection. Considerable spectral broadening is observed already at 5~\textmu W of injected power. Due to the small power required for achieving a tight lock, a single line isolated from the comb spectrum provides enough power to act as a Master oscillator. In our work, a pair of etalons was employed as tunable optical filter. While this solution is based on free-space optics, the presented method is very general and will greatly benefit from photonic integration. In particular, optical filters based on microresonators\cite{komljenovic2017widely} could enable the photonic integration of a fully phase-stabilized dual-comb spectrometer operating in the molecular fingerprint region. Optical synchronization of the dual-comb spectrometer has been shown to provide several benefits including a significantly increased SNR and a periodic interferogram. While the latter property may seem subtle, it greatly reduces the complexity of data acquisition and analysis. Firstly, the amount of data is decreased by several orders of magnitude thanks to coherent averaging, while no spectral information is lost. Secondly, the periodicity of the signal allows to employ sub-sampling techniques\cite{sterczewski2020subsampling}. As a consequence, slower and therefore significantly cheaper digitizers can be used to acquire the dual-comb spectra. Finally, optical injection locking will enable coherent and high-resolution spectroscopy based on spectral interleaving~\cite{gianella2020high}, while the frequency axis can be calibrated by comparing a single line to an optical reference~\cite{komagata2021all}.

\section*{Methods}

\subsection*{RF-packaged QCL frequency combs}
The QCL frequency combs used in this work are based on a plasmon-enhanced waveguide and emit a power of roughly 200 mW in continuous wave at 273 K. The lasers were packaged with a printed circuit board (PCB) optimized for RF injection and extraction. One port of the PCB connected to the rear end of the QCL cavity allows for coherent electrical injection locking of $f_{rep}$. A Windfreak SynthHD Pro synthesizer delivers the injection signals for both combs. A second port is connected to the front end of the cavity and used to monitor the beatnotes inside the QCL. A third port delivers the DC laser current.

\subsection*{Wafer-based Fabry-P\'er\^ot interferometers}

The optical Vernier filter used in this work consists of two etalons. The latter are based on two doubleside-polished Ge and GaAs wafers. High-reflectivity coatings are deposited on both sides of the wafers using ion beam sputtering. These coatings form the mirrors of the Fabry-P\'er\^ot cavity.
\linebreak

\section{Funding}
This work was supported by the BRIDGE program of the Swiss National Science Foundation and Innosuisse under the Project CombTrace (No.40B2-0-176584) and the Qombs Project by the European Union's Horizon 2020 research and innovation program (No. 820419). B.S. and N.O. have received funding from the European Research Council (ERC) under the European Union's Horizon 2020 research and innovation programme (Grant agreement No. 853014).

\section{Acknowledgments}
The authors thank Dr. Matthew Singleton for fruitful discussions.

\bibliographystyle{naturemag}
\bibliography{literature}
\end{document}